\RequirePackage{fix-cm}
\documentclass[smallextended]{svjour3}       
\smartqed  
\usepackage{graphicx}
\graphicspath{{/Images/}}
\begin{document}

\title{Niobium nitride thin films for very low temperature resistive thermometry  
}

\author{Tuyen Nguyen  \and Adib Tavakoli \and Sebastien Triqueneaux \and Rahul Swami \and Aki Ruhtinas \and Jeremy Gradel \and Pablo Garcia-Campos \and  Klaus Hasselbach \and Aviad Frydman \and Benjamin Piot \and Mathieu Gibert \and Eddy Collin \and Olivier Bourgeois
}


\institute{T. Nguyen, A. Tavakoli, S. Triqueneaux, R. Swami, J. Gradel, P. Garcia-Campos, K. Hasselbach, B. Piot, M. Gibert, E. Collin, O. Bourgeois \at
	Univ. Grenoble Alpes, Institut NEEL, CNRS-25 avenue des Martyrs, F-38042 Grenoble, France \\
	Tel.: +334 76 88 12 17 \\
	\email{olivier.bourgeois@neel.cnrs.fr}  
	\and
	A. Ruhtinas \at
	Nanoscience Center, Department of Physics, University of Jyv\"{a}skyl\"{a}, PO Box 35, FI-40014 Finland
	\and
	A. Frydman \at
	The Department of Physics, Bar Ilan University, Ramat Gan 52900, Israel
}

\date{Received: date / Accepted: date}

\maketitle

\begin{abstract}

We investigate thin film resistive thermometry based on metal-to-insulator-transition (niobium nitride) materials down to very low temperature. The variation of the NbN thermometer resistance have calibrated versus temperature and magnetic field. High sensitivity in tempertaure variation detection is demonstrated through efficient temperature coefficient of resistance. The nitrogen content of the niobium nitride thin films can be tuned to adjust the optimal working temperature range. In the present experiment, we show the versatility of the NbN thin film technology through applications in very different low temperature use-cases. We demonstrate that thin film resistive thermometry can be extended to temperatures below 30~mK with low electrical impedance.

\keywords{resistive thermometry \and niobium nitride \and thin film \and nanoscale }
\end{abstract}

\section{Introduction}
\label{intro}

For applications in cryogenics, it becomes more and more useful to cover large temperature ranges with a single thermometry. Most of the conventional thermometers like platinum for high temperature or germanium for low temperature do not allow measurements between ambient and very low temperature ($<$~50~mK) with the same sensor.
Moreover, measuring with a high sensitivity low temperature is of crucial importance for monitoring cryogenic systems, high magnetic field coils, space satellite applications, bolometry, investigating thermal physics or characterizing materials below 1~K \cite{Giazotto,Enss2008,Pickett}.

The current existing resistive thermometry such as Ge crystal, ruthenium oxide, based on zirconium oxynitride (Cernox) etc... may have intrinsic limitations since they are a not all versatile, for most of them cannot be downscaled and have high resistance at the lowest temperature (above 100~kOhm below 1~K) \cite{Schuster1994,Rubin1997,Fisher2005,Querlioz2005,Tagliati2012,Olivieri2015,Swinehart,Lin}. 

Consequently, being capable of measuring temperatures from 300~K to below 100~mK in various kinds of applications remains a challenge since commercial solutions cannot cover this large temperature range \cite{Olivieri2015}. Regular doped semiconducting materials like silicon or germanium have a resistance too high to be measured at low temperature when prepared as thin films, preventing them from use as efficient temperature sensors especially if downscaling is required. 

Among potential transducers, a particular kind of resistive materials can be of great interest for low temperature thermometry: the metal-to-insulator transition (MIT) materials \cite{MIT}. Indeed, either due to Coulomb repulsion or to disorder, some systems may exhibit a significant increase of resistance as the temperature is lowered as a consequence of electron localization. This has been theoretically introduced by Anderson in 1958 and Mott in 1968 through different physical mechanisms leading to resistance variation having similar power law in temperature.

As for Anderson case of MIT, a quantum phase transition is expected at $T=0$ where the transition to the insulating state is driven by disorder. This can be obtained in numerous materials prepared as thin films or ultra-thin films like InO$_{\rm x}$, NbSi, amorphous metal, TiN \cite{Baturina2007,Ovadia2009,Crauste2011,amorphmetal,Couedo2016}. Here, we present a thermometry based on niobium nitride (NbN) thin film taking benefit of the significant increase of resistance as the temperature is lowered. It allows measurements of temperatures from 300~K down to 30~mK (three orders of magnitude) \cite{Bourgeois2006}, already widely used for very sensitive thermal measurements like thermal conductance \cite{Bourgeois2007,Heron2009a,Wingert2012,Sikora2012,Sikora2013,Blanc2013,Tavakoli2018} or heat capacity experiments \cite{Ong2007,Lopeandia2010,Souche2013,Poran2017}. We demonstrate that it provides low impedance thermometer and high sensitivity in temperature (above 1~K$^{-1}$) along with a moderate magnetoresistance. The thin film technology allows the deposition of NbN on many different supports (bulk silicon, nanowires, membranes, fibers etc...) showing the high versatility of this thermometry.

\section{Thermometer deposition}
\label{sec:1}

The deposition of NbN$_{x}$ is made by magnetron sputtering of a pure Nb target (99.95 \%) in an Ar/N$_2$ gas mixture of high purities (99.99 \%).  The deposition chamber is maintained under vacuum with a turbo-molecular pump and a primary pump along with a liquid nitrogen trap (the pressure has to be less than $5\times10^{-7}$~mbar). Mass flow controllers precisely regulate the amount of gas injected into the deposition chamber, a butterfly valve allowing a fixed work pressure of $2\times10^{-2}$~mbar.

\begin{figure*}
	\includegraphics[width=1.0\textwidth]{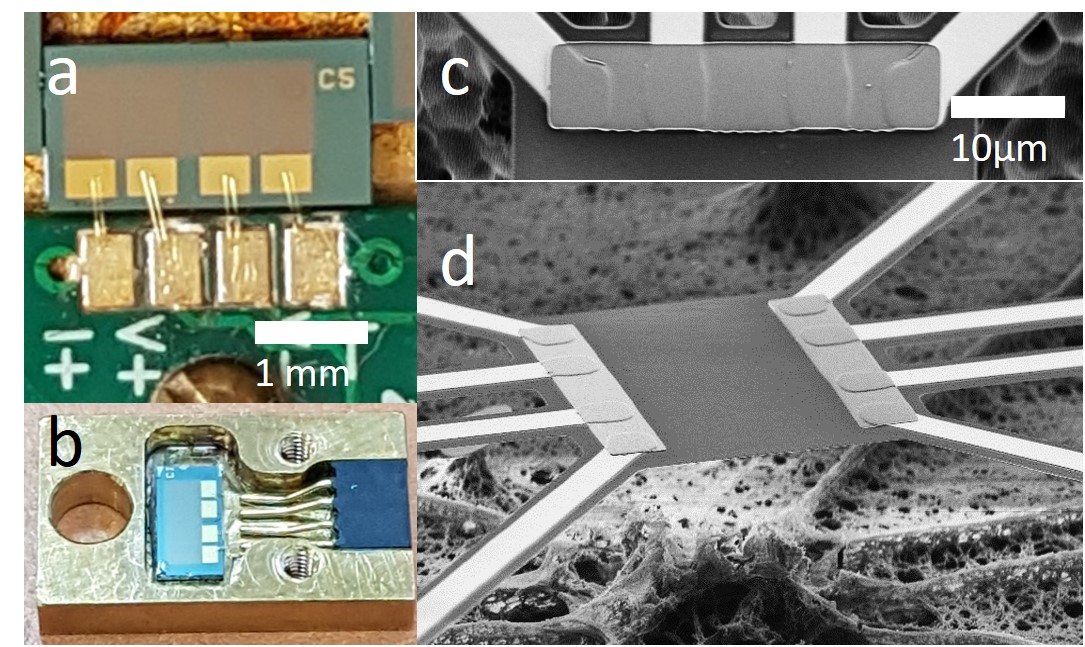}
	\caption{Two examples of NbN thin film thermometers at the millimeter or micrometer scale. a) NbN thermometer deposited on Si/SiN substrate. b) NbN on Si/SiN installed in an EM shield. c) Top view of a NbN thermometer in SiN membrane. d) Example of a suspended sensor functionalized with NbN thermometer for low temperature thermal studies. }
	\label{fig:1}       
\end{figure*}

The pulsed DC power supply allows electrons to follow helical paths around magnetic field lines undergoing more ionizing collisions with the neutral elements at the surface of the target. This forms an electric current with both AC and DC components. During their flight between the target and the substrate, niobium atoms react with the nitrogen ions of the plasma and form (niobium) nitride. This is essential because the nitrogen contents of the plasma will determine the nitrogen content of the final NbN layers. The variations of the resistance of this material as a function of temperature are very sensitive to the nitrogen content in the layers.

The sputtering parameters as the gas mixture ratio, the frequency and the power supplied of the magnetron are carefully controlled to adjust the nitrogen content in the layer and achieve the good stoichiometric composition and microstructure. Depending on the ratio $x =$N/Nb, the electrical properties are very different. For $x < 0.6$, the material behaves like a superconductor with a critical temperature of about 15~K. For $x > 1.1$, the material behaves as a Mott-Anderson insulator characterized by a MIT transition at a temperature dependent on the nitrogen content of the layer. A material with a high nitrogen concentration like NbN$_{1.7}$ will give a thermometer having a metal-insulator transition at high temperature. Reversely, thin layers of NbN$_{1.6}$ will be perfectly adapted to low temperature thermometry (for more details on the composition analysis by Rutherford Back Scattering see Ref.~\cite{Bourgeois2006}). This gives a short window over which tuning of nitrogen content can be performed for optimizing the performance of the thermometers. Finally, all thin films are annealed at 150 Celsius degrees for fifteen hours. This helps to stabilize the film avoiding drift and get better reproducibility under thermal cycling.

In order to precisely adjust this crucial Ar/N$_2$ ratio, thin film of NbN are deposited on sapphires as test samples whose electrical resistance is measured between 77~K and 300~K. This electrical resistance ratio, called the resistive ratio $RR$ ($RR=R_{77\rm {K}}/R_{300 \rm {K}})$) is used to qualify the expected properties of the NbN thin films. We generally deposit a layer of niobium nitride of about 70~nm using the same deposition parameters as those set during the test measurements done on sapphire. The layer thickness is monitored by the deposition time previously calibrated on test samples.

Fig.~\ref{fig:1} shows the versatility of the NbN deposition; two examples of thermometers are given, one on the millimetric size with deposition done on silicon substrate (caped with 100~nm of SiN for electrical isolation), the other on the micrometer size with a NbN thermometer included on a SiN membrane calorimetric suspended sensor \cite{Tavakoli2018}.

\section{Thermometer calibration}
\label{sec:2}

The NbN thermometers are installed in the vacuum chamber of a dilution cryostat. They are shielded against high frequency variation of the electromagnetic (EM) field and all the electrical leads are filtering EM radiations above 50~kHz using passive filters at room temperature. All the thermometers have been calibrated on a speer carbon resistance, itself calibrated on a SRD1000 (having superconducting fixed points) and on a Coulomb Blockade primary thermometer \cite{SRD,CBT1,CBT2}. 

\begin{figure}
	\includegraphics[width=1.0\textwidth]{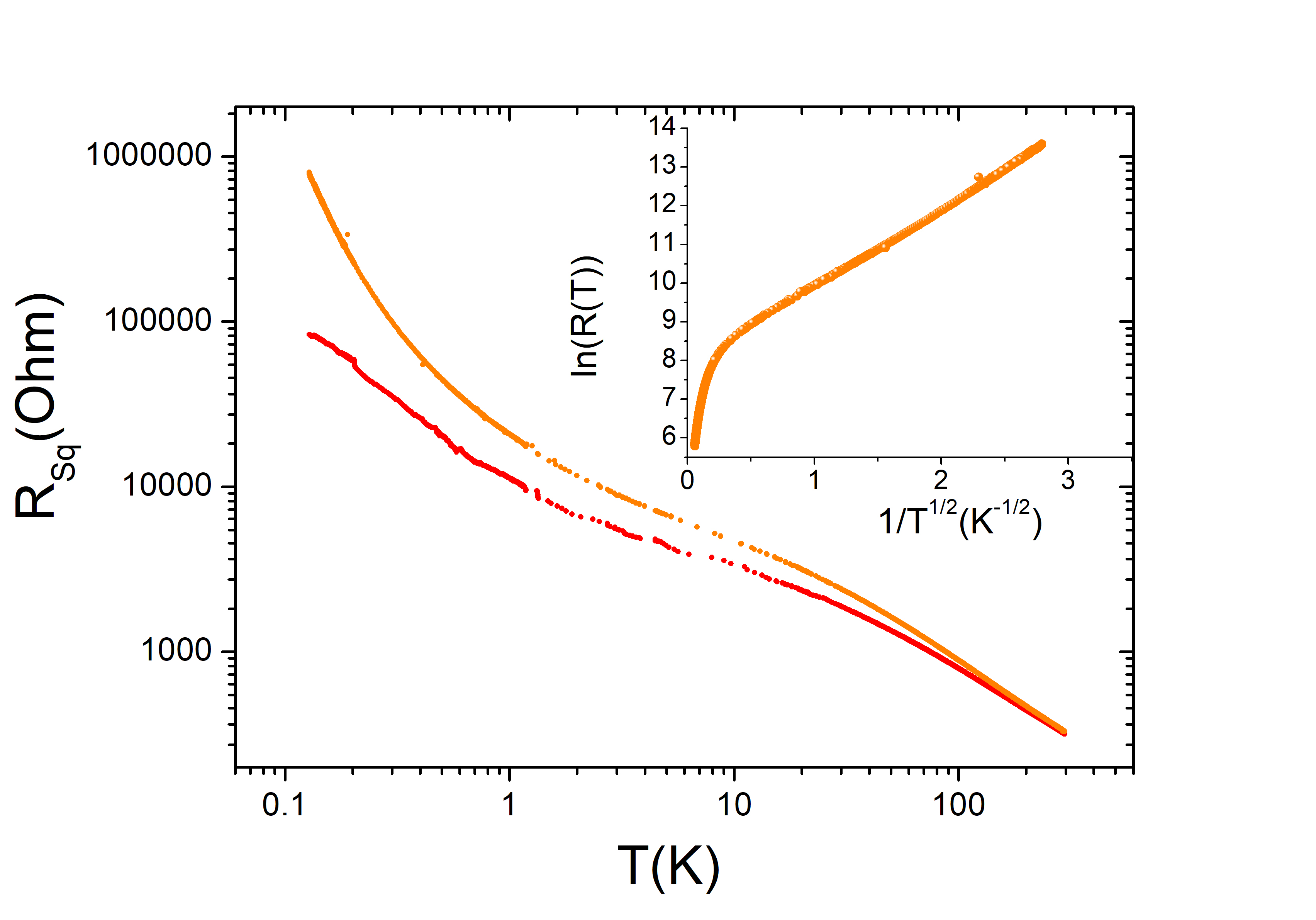}
	\caption{Resistance per square versus temperature measured from room temperature down to 100~mK on two NbN thin films having been prepared with two different resistive ratio ($RR=R_{77\rm {K}}/R_{300 \rm {K}})$). The orange dots corresponds to $RR=5$ and the red dots to $RR=4$. In inset, ln(R(T)) as a function of $1/T^{1/2}$ showing the linear behavior for temperature below 30~K.}
	\label{fig:2}       
\end{figure}

Large thermometers deposited on silicon substrates (see Fig.~\ref{fig:1}~a and b) have been first calibrated down to 100~mK. Two kinds of thermometers prepared with different nitrogen contents have been considered giving two resistive ratio $RR$ of 4 and 5. The calibration of the resistance per square versus temperature is presented in Fig.~\ref{fig:2}; the resistance per square ($R_{Sq}$) corresponds to the resistance of a film of thickness $e$ and identical width and length. The significant increase of resistance at low $T$ demonstrates the capacity of NbN thermometers to measure temperatures over a very broad range from room temperature to 100~mK. It can be noticed that a lower resistive ratio will give a lower electrical impedance at the lowest temperature. This will be used to adjust resistance for very low temperature measurement. 

As for the temperature dependence of the resistance, it departs from a regular Mott law $R=R_0 exp(T_0/T)^{1/4}$ as it has been observed in the past for higher temperature \cite{Bourgeois2006}. The actual variation is closer to a variable range hoping (Efros-Shklovskii) prediction of $R=R_0 exp(T_0/T)^{1/2}$ as it can be seen in the inset of Fig.~\ref{fig:2}.

\begin{figure}
	\includegraphics[width=1.0\textwidth]{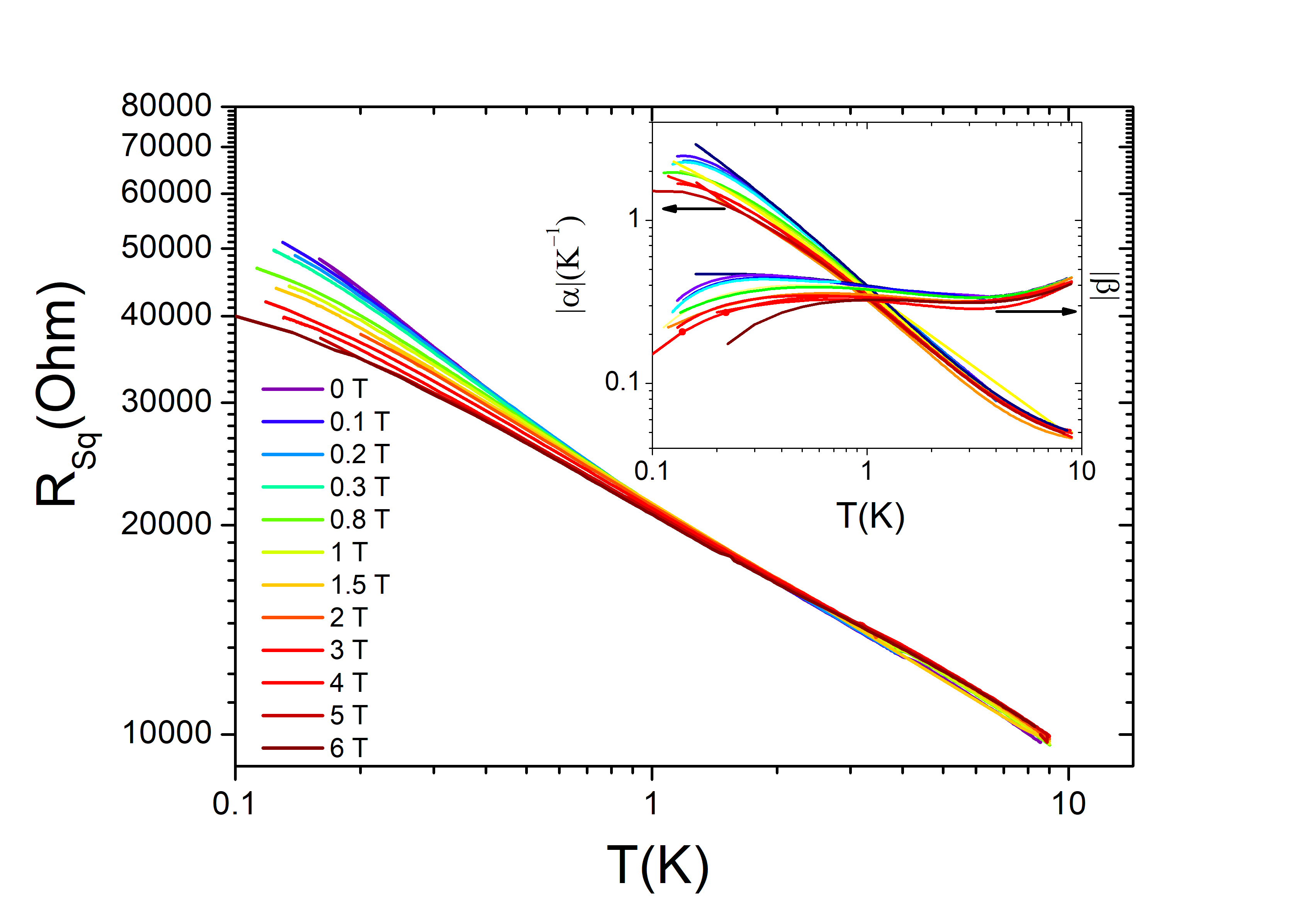}
	\caption{Resistance per square versus temperature measured from 10~K down to 100~mK under various magnetic field, up to 6~Tesla, on a NbN thermometer of Fig.~\ref{fig:1} prepared to have a resistive ratio of $RR=3$. In inset, the absolute value of the temperature coefficient of resistance $|\alpha|=\frac{1}{R} \frac{dR}{dT}$  and the dimensionless parameter $|\beta|=\frac{T}{R} \frac{dR}{dT}$ are plotted as a function of temperature.}
	\label{fig:3}       
\end{figure}

For very low temperature measurements, a thermometer of $RR=3$ has been prepared. The calibration of the thermometer under magnetic field has been performed up to 6~Tesla (see Fig.~\ref{fig:3}). As very low magnetoresistance is evidenced above 0.8~K (less than 5\% at 6~T), variation of resistance up to 20\% at 150~mK is observed under 6~Tesla, a value comparable to the one observed with ruthenium oxide thermometer (RuO$_2$). If this does not prevent from using NbN for measurements under magnetic field, however, this has to be taken into account for corrections especially at the lowest temperature \cite{fortune,zhang}.

The temperature coefficient of resistance calculated through $|\alpha|=\frac{1}{R} \frac{dR}{dT}$  (shown in inset of Fig.~\ref{fig:3}) can be bigger than $1$~K$^{-1}$, a particularly high value for thin film technology. The thermometer performance can be also characterized by the dimensionless parameter $|\beta|=\frac{T}{R} \frac{dR}{dT}$, better than $|\alpha|$ for comparison with other thermometry. As shown in inset of Fig.~\ref{fig:3}, the $|\beta|$ coefficient is rather constant over all the temperature range of the measurement around a value of 0.5. This means that for a resistance of 5~kOhm, using a bias current of 1~nA, a variation of temperature as small as 20~$\mu$K can be detected at $T=0.1$~K by averaging over one minute. This sensitivity is limited by the intrinsic noise of the preamp which is on the order of 0.6~nV$\sqrt{\rm Hz}^{-1}$.

If the thermalization of the large NbN thermometer on bulk Si substrate is not questioned, a more thorough thermal balance analysis has to be done for small size thermometer, as shown in Fig.~\ref{fig:1}~c and d.

Very low current source are commonly used to minimize thermometer self-heating. For each temperatures, linearity of I-V characteristic have been checked ruling out any potential over-heating, at 50~mK this has been done for current ranging from 20~pA to 1~nA. If we consider a thermometer of 1~kOhm and a measuring current of 0.5~nA, this implies a power dissipated in the electron bath of the thermometer of $P=2.5\times 10^{-16}$~Watt. 


This power can be used to estimate the temperature difference between the electron bath and the phonon bath of the NbN film as well as the temperature gradient between the phonons of the NbN and the phonons of Si or SiN; for this, we consider only the small size NbN film of the following geometry: 70~nm thick, for a surface of 6~$\mu$m by 40~$\mu$m as presented in Fig.~\ref{fig:1}~c. Regarding the possible electron-phonon decoupling, we use the well-known electron-phonon coupling relation:
\begin{equation}
\frac{P}{V}=g_{e-ph}\left(T^5_{e-/NbN}-T^5_{ph/NbN}\right)  
\end{equation}

where $P$ is the power dissipated in the electron bath, $V$ the volume of the thermometer, $T_{e-/NbN}$ and $T_{ph/NbN}$ the temperature respectively of the electron and phonon bath of the thermometer, and $g_{e-ph}$ the electron-phonon coupling constant which is of the order of $10^{2}$~W.K$^{-5}$.cm$^{-3}$ (we take the same coupling constant as in amorphous NbSi, a very similar material \cite{Humbert}). By using the known dimensions of the NbN thermometer, we can estimate the temperature of the electrons. Indeed if $T_{ph/NbN}$ is regulated at 50~mK then $T_{e-/NbN}=53$~mK for a measuring current of 0.5~nA. By working with smaller current than 0.5~nA no over-heating will be expected.

Regarding the coupling between the phonon baths in the NbN and in the silicon substrate, a rough estimate of the Kapitza thermal conductance indicates an expected thermal conductance of the solid-solid contact between NbN and Si or SiN of $K_{Kapitza}=1.4\times10^{-10}$~W.K$^{-1}$ \cite{Swartz}. This analysis shows that micrometer size thermometers can be prepared and used for very low temperature physics under the careful choice of the measuring currents \cite{Souche2013,Tavakoli2018}.

\begin{figure}[tb]
	\includegraphics[width=1.0\textwidth]{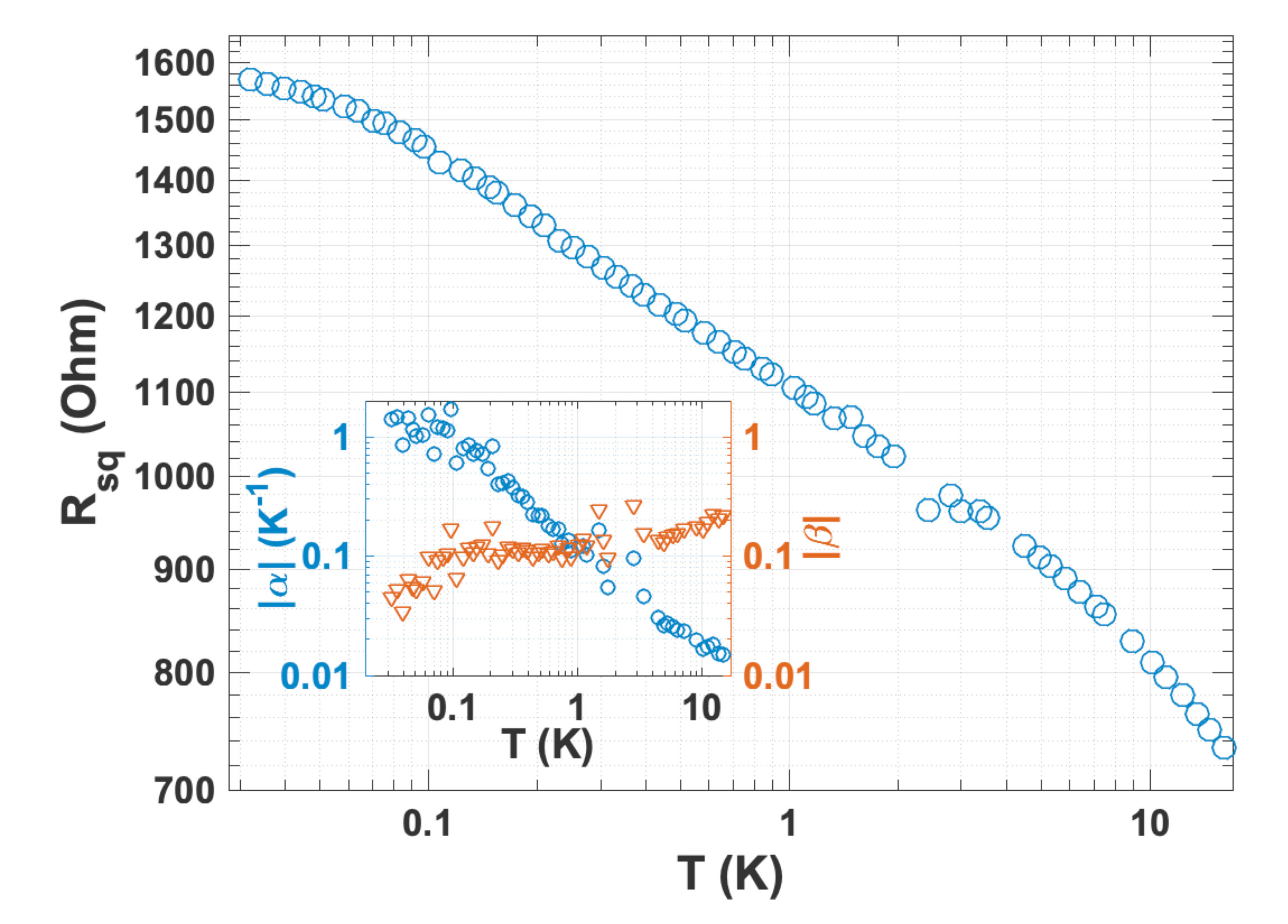}
	\caption{Resistance per square versus temperature measured below 10~K (down to 30~mK) on a low resistance NbN thin film having a resistive ratio of $RR=2.7$. In inset, the absolute value of the temperature coefficient of resistance $|\alpha|=\frac{1}{R} \frac{dR}{dT}$  and the dimensionless parameter $|\beta|=\frac{T}{R} \frac{dR}{dT}$ are plotted as a function of temperature.}
	\label{fig:4}       
\end{figure}

\section{Very low temperature limit}
\label{sec:3}
This part is dedicated to the test of NbN thermometry for temperature below 100~mK. To be adapted to the very low temperature, the $RR$ has been decreased to 2.7. The measurement has been done in a dedicated dilution fridge. 

The sensor has been calibrated against a combined SRD1000-CMN1000 system \cite{leiden}. The presence of eight fixed points on the SRD1000 device between 30~mK and 1.2~K allows a fine calibration of the CMN1000 over the full range. Above 1.2~K, a Cernox type sensor calibrated by LakeShore has been used. We have checked the calibration against the upper temperature superconducting fixed points (resp. 3411, 4987 and 7191~mK for In, Va and Pb). The resistance is determined via a four wire measurement using a resistance bridge dedicated to low temperatures \cite{MMR3}. Excitation current below 1~nA has been used. 

The calibration in temperature down to 30~mK is shown in Fig.~\ref{fig:4}. This shows that this type of thermometry is particularly adapted to very low temperature with a temperature coefficient $|\alpha|$ of 1~K$^{-1}$, and the dimensionless parameter $|\beta|$ of 0.5. Even though a slight deviation from linearity in the log-log plot is limiting the sensitivity, it is validating the appropriateness of the NbN for measuring temperature below 50~mK, with an expected change of resistance of 2~Ohm per milliKelvin. Further study will be required to address this deviation in resistance at the lowest temperature. This could be due to not good enough filtering of electromagnetic radiations, slight over-heating or intrinsic resistance variation of MIT materials as the temperature approaches the absolute zero. Clearly, the preparation of the thermometer for very low temperature is the result of a trade-off between a low thermometer impedance but still a high sensitivity.


\section{Conclusions}
\label{concl}

This work has been devoted to show the ability of NbN thermometry for low and very low temperature measurements. The versatility of NbN thermometers has been evidenced through the preparation of thermometers from millimeter to micrometer sizes. Through the possible change of nitrogen content in the thin films, a broad working temperature range can be obtained. Temperatures from above 300~K down to 30~mK, four orders of magnitude, have been measured with the same NbN transducer thanks to the continuous exponential decrease of resistance as the temperature is increased. A very high sensitivity has been shown down to 30~mK, allowing the detection of temperature changes as small as 20~$\mu$K. It is also highlighted that measuring very low temperatures can be addressed with the same thin film technology by adapting the nitrogen content. This is demonstrating all the capabilities of NbN thermometers that can be deposited on various dielectric support and adapted to any cryostat or downscaled sensor that requires the precise management of temperature over a very broad range.

\begin{acknowledgements}
	
We thank the micro and nanofabrication facilities of Institut N\'eel CNRS: the P\^ole Capteurs Thermom\'etriques et Calorim\'etrie, Nanofab for their help in the preparation of the samples and the experiments. We have also benefited from the support of the Pole Cryogenie and Pole Electronique. The research leading to these results has received funding from the European Union's Horizon 2020 Research and Innovation Programme, under grant agreement No.~824109, the European Microkelvin Platform (EMP), the EU project MERGING grant No.~309150, ERC CoG grant ULT-NEMS No.~647917, the authors also acknowledges the financial support from the ANR project QNM Grant No.~040401,  the Laboratoire d'excellence LANEF in Grenoble (ANR-10-LABX-51-01), the ANR project Tiptop ANR-16-CE09-0023; PDG aknwoledges funding from Innovation Programme under the Marie Sk\l{}odowska-Curie grant agreement No.~754303 and the Fondation des Nanosciences (FCSN 2018~02D), and AR from Erasmus EU programme.

\end{acknowledgements}


\begin{thebibliography}{}
\bibitem{MMR3}
The MMR3 resistance bridge has been developed at the Institut N\'eel and is now licenced to Cryoconcept.
\end{thebibliography}


\end{document}